# Fractal Dimensions Derived from Spatial Allometric Scaling of Urban Form

Yanguang Chen, Yihan Wang, Xijing Li

(Department of Geography, College of Urban and Environmental Sciences, Peking University, Beijing 100871, P.R.China. E-mail: chenyg@pku.edu.cn)

**Abstract**: The improved city clustering algorithm can be used to identify urban boundaries on a digital map, and the results are a set of isolines. The relationships between the urban measurements within the variable boundaries follow allometric scaling law, which indicates spatial allometry of cities. This paper is devoted to exploring the fractal dimension proceeding from urban spatial allometry. By theoretical reasoning and empirical analysis of urban traffic network, we can derive a set of fractal dimension from the spatial allometry and reveal the basic property of the new fractal parameters. The findings are as follows. First, the fractal dimension values of traffic lines are higher than those of traffic nodes. Second, the fractal dimension values based on variable boundaries are lower than those based on the concentric circles. Conclusions can be reached that the fractal dimensions coming from spatial allometry are a type of correlation dimension rather than capacity dimension, and the relative growth rate of traffic points is greater than that of traffic nodes. This study provides new way of understanding allometry, fractals, and scaling in urban systems.

**Key words**: allometric scaling; city clustering algorithm; fractals; urban form and growth; urban boundary; traffic network

## 1 Introduction

One of important power-law behaviors of urban system is allometric growth, and allometry in cities seems to be a scaling pattern emerging from urban evolution. A city is a kind of self-organized system (Allen, 1997; Haken and Portugali, 1995; Portugali, 2000), which can be treated as typical complex adaptive system (Holland, 1995). The process of self-organizing evolution of cities follows the law of allometric scaling, which has been researched by many scientists (Arcaute *et al*, 2005; Batty, 2008; Batty, 2013; Batty and Longley, 1994; Bettencourt, 2013; Bettencourt *et al*, 2007; Chen, 2014a; Chen and Zhou, 2008; Louf and Barthelemy, 2014a; Louf and Barthelemy, 2014b; West, 2017; Luo and Chen, 2014). Allometric modeling can be utilized to describe the scaling relations between urban and rural population (Chen, 2014b; Naroll and Bertalanffy, 1956), relations between an urban system and its central city (Beckmann, 1958; Zhou, 1995), relations between urban area and population size (Batty and Longley, 1994; Lee, 1989; Lo and Welch, 1977; Nordbeck, 1971), relations between urban area and boundary (Batty and Longley, 1994; Chen, 2013), relations between different cities of an urban system (Chen and Jiang, 2009), scaling of building geometries (Batty *et al*, 2008; Gould, 1971), and so on. Among all these models, the most frequent one is the urban area-population allometry, which indicates the scaling relation between size and shape in the growth of human communities. The common area-population allometry falls into two types: *longitudinal allometry* and *transversal allometry* (Pumain and Moriconi-Ebrard, 1997). The former is on urban growth and can be fitted to a pair of time series of a given city, while the latter is on the rank-size distributions of cities and can be fitted to a pair of datasets of an urban system at a given time. The longitudinal allometry represents a temporal scaling, while the cross-sectional allometry suggests a hierarchical scaling.

In theory, there should also be the third allometric scaling relation between urban area and population, namely, spatial allometry. According to the ergodic hypothesis, a temporal process of a geography system not only corresponds to a size distribution (Batty and Longley, 1994), but also corresponds to a spatial distribution (Harvey, 1969). By means of a digital map, we can draw a set of isograms of urban density. Each isoline gives an urban area and an urban population value. If urban form follows the law of allometric scaling, we can find a power-law relation between area and population based on the isograms. A discovery is that the intraurban patterns follow a set of spatial allometric scaling laws, which can be verified with the observed data based on the variable boundaries of cities. By ArcGIS technology, we can identify urban boundary using spatial search method. Changing searching radius of a city defined in a 2-dimensional space, we have different urban boundary lines. Different urban boundaries include different urban areas, total length of streets, and node number of traffic networks. Thus three power laws can be found to describe the scaling relationships between the points (nodes), lines (streets), and area (regions), which represent the basic spatial elements of geographical systems. These power laws compose the basic models of spatial allometry of urban growth and form.

An allometric relation is in fact a fractal measure relation between size and shape. According to the principle of dimension consistency (Lee, 1989; Mandelbrot, 1983; Takayasu, 1990), one measure is proportional to another measure if and only if the two measures share the same dimension in a mathematical space (Chen, 2015). Otherwise, the proportion comes into being if the dimensions of two measures are transformed into the identical value by extracting certain roots. This leads to the power law relations indicative of spatio-temporal scaling and fractal dimension. An allometric scaling exponent proved to the ratio of one fractal dimension to another fractal dimension (Chen, 2010). However, what is the property of the fractal dimensions derived from spatial allometry? How to understand the relationship between the allometry-based fractal dimension and the conventional radial fractal dimension of cities? This paper is devoted to researching the fractal parameters based on spatial allometry of urban morphology. The rest of the article is organized as follows. In Section 2, a set of models on spatial allometry of urban form and the related fractal dimension models are presented, and the relationships between the scaling exponents and fractal dimension is clarified. In Section 3, empirical analyses on spatial allometry and radial fractal dimension are made by means of the observational data of 10 Chinese cities. In Section 4, the related questions are discussed, and the association of the spatial allometry and fractal dimensions with space types is illustrated. Finally, the discussion is concluded by summarizing the main points of this work.

# 2 Models

## 2.1 Fractality and variable urban envelopes

Spatial allometry can be associated with fractal cities, and allometric models are in essence fractal models. A great number of empirical studies and theoretical analyses showed that urban form bears fractal properties (Batty and Longley, 1994; Benguigui *et al*, 2000; Chen and Wang, 2013; Feng and Chen, 2010; Frankhauser, 1994; Jiang and Yin, 2014; Makse *et al*, 1995; Makse *et al*, 1998; Thomas *et al*, 2010). Fractals indicate scaling, and fractal cities are free of characteristic scale. Therefore, it is impossible to find an accurate and objective boundary for a city. Urban boundary depends on the way of spatial measurements. In urban geography, there exist three key concepts of cities: city proper (CP), urban agglomerations (UA), and metropolitan areas (MA) (Davis, 1978). The second one is sometimes replaced by urbanized area (UA). Different city concepts agree with different urban boundaries. The boundary curves of cities are termed *urban envelopes* (Batty and Longley, 1994; Longley *et al*, 1998). Today, there are at least three scientific approaches to determining urban envelopes for cities. The first is the city clustering algorithm (CCA) proposed by Rozenfeld *et al* (2008, 2011), the second is the fractal-based method presented by Tannier *et al* (2011), and the third is to derive what are called 'natural cities' by clustering street nodes/blocks (Jiang and Jia, 2011). An urban boundary determined by one of these three methods corresponds to an urban agglomeration or urbanized area.

Urban density follows the law of distance decay, which reads that the density of population, land use, and streets and roads, and so on, decreases from the center to the fringe of a city. Urban population density follows Clark's law and takes on exponential decay (Clark, 1951), and urban road density follows Smeed's law and takes on power-law decay (Smeed, 1963). The density of urban land use is complex. The land-use density of some cities follows inverse power law, while other cities follows negative exponent law. Both the inverse power law and negative exponential law belong to the distance-decay laws. Due to distance decay of urban density, the method of spatial search can be employed to determine urban boundary (Jiang and Jia, 2011; Rozenfeld *et al*, 2008; Tannier *et al*, 2011). In a digital map, a city cluster is in fact a set of pixels. Writing an automated search program, we can make a spatial search from city core to periphery with given searching radius. If the distance between two adjacent pixels is less than the searching radius, the search process will continue, otherwise the search process will be confined and cannot extend to outside. Finally, the spatial search will yield a boundary curve, which is just the urban envelope based on given searching radius. Obviously, changing the searching radius, we will have different urban envelopes, which indicates variable urban boundary. If we can find the characteristic length of the searching radius, we can define an objective urban boundary, but this problem goes beyond the task of this study and remains to be solved in a companion article.

## 2.2 Allometric scaling in intraurban patterns

Allometric scaling can be found in the spatial pattern of a city based on variable boundaries. As soon as a boundary is figured out, we will have at least three results of spatial measurements: urbanized area within the boundary (plane), total length of streets (lines), and number of street nodes (points). For short, the three measurements can be called *urban area*, *street length*, and *node number*, which represent plane, line, and point, respectively. However, as indicated above, an urban boundary depends on the length of searching radius. Different radius yields different urban boundary, inside which the urban area, street length, and node number will be different. Changing the searching radius again and again, we will have a series of results, including a dataset of urban areas, a dataset of street lengths, and a dataset of node numbers. The observational datasets show that the three spatial measurements increase exponentially with the increase of searching radius. Thus we have

$$A(s) = A_0 e^{as}, \tag{1}$$

$$L(s) = L_0 e^{cs}, \tag{2}$$

$$N(s) = N_0 e^{ks}, \tag{3}$$

where $s$ denotes the length of searching radius, $A(s)$ refers to urban area, $L(s)$ to street length, and $N(s)$ to node number, $A_0$, $L_0$, $N_0$, $a$, $c$, and $k$ are parameters. From equations (1), (2), and (3) it follows three power-law relations such as

$$L(s) = \mu A(s)^b, \tag{4}$$

$$N(s) = \eta A(s)^v, \tag{5}$$

$$N(s) = \xi L(s)^\sigma, \tag{6}$$

in which the powers $b=c/a$, $v=k/a$, $\sigma=k/c=v/b$ denotes scaling exponents, and $\mu=L_0A_0^{-b}$, $\eta=N_0A_0^{-v}$, and $\xi=N_0L_0^{-\sigma}$ are proportionality coefficients. Obviously, equation (6) can be derived from equations (4) and (5).

A scaling relation of urban form is always associated with fractal structure. The dimension of urban area can be regarded as $D_a=d=2$. Suppose that the fractal dimension of network of streets is $D_s$, and the fractal dimension of node distribution is $D_n$. According to the fractal measure relation (Chen, 2010; Feder, 1988; Mandelbrot, 1983; Takayasu, 1990), the relationships between the scaling exponents and fractal dimensions are as follows

$$b = \frac{D_s}{D_a} = \frac{D_s}{2}, \tag{7}$$

$$v = \frac{D_{\mathrm{n}}}{D_{\mathrm{a}}} = \frac{D_{\mathrm{n}}}{2}, \tag{8}$$

$$\sigma = \frac{D_{\mathrm{n}}}{D_{\mathrm{s}}} = \frac{v}{b}. \tag{9}$$

Among the three fractal parameter equations, one of them can be derived from other two ones. For example, equation (9) can be derived from equations (7) and (8). Using these parameter equations, we can estimate the fractal dimensions and the ratio of different fractal dimensions of urban form indirectly.

## 2.3 Spatial allometry and radial fractal dimension

The fractal parameters can be understood through a special spatial allometry of urban growth and form. In fact, the allometric scaling models can be built in the perspective of standard circle, and thus the allometry can be directly associated with fractal models. For the urban area $A(s)$ within a boundary curve (Figure 1a), the urban envelope corresponds to a circle of equal area $A(r)$ (Figure 2b), that is, we have

$$A(s) = A(r) = \pi r^2, \tag{10}$$

where $r$ denotes a radius or a distance from a city center, $\pi$ is the circumference ratio. This indicates that $A(r)$ is the circular area measured by the given radius $r$. The area in the largest circle, $A(R)$, represents urban field, here $R=r_{\max}$, and $F=2R$ is termed Feret's diameter (Batty and Longley, 1994; Longley *et al*, 1991). Thus, equation (4) can be rewritten as

$$L(r) \propto A(r)^{D_{\mathrm{s}}/D_{\mathrm{a}}} \propto L_1 r^{D_{\mathrm{s}}}, \tag{11}$$

where $L(r)$ denotes the total length of streets within a radius of $r$ of the city center, "$\propto$" means "be proportional to", and $L_1$ is a proportionality coefficient. Accordingly, equation (5) can be re-expressed as

$$N(r) \propto A(r)^{D_{\mathrm{n}}/D_{\mathrm{a}}} \propto N_1 r^{D_{\mathrm{n}}}, \tag{12}$$

where $N(r)$ represents the number of street nodes within a radius of $r$ from the city center, and $N_1$ is a proportionality coefficient. From equations (11) and (12) it follows

$$N(r) \propto L(r)^{D_{\mathrm{n}}/D_{\mathrm{s}}}, \tag{13}$$

which is just the allometric relation corresponding to equation (6). Equations (11), (12), and (13)

can be treated as special cases of equations (4), (5), and (6), respectively. Note that the city radius $r$ differs from the above-mentioned searching radius $s$. They are two different concepts in the context. Taking derivatives of $A(r)$ in equation (10) and $L(r)$ in equation (11) with respect to $r$ yields a density function of transport networks such as

$$\rho_{\mathrm{L}}(r)=\frac{\mathrm{d}L(r)}{\mathrm{d}A(r)}=\frac{D_{\mathrm{s}}L_1 r^{D_{\mathrm{s}}-1}}{2\pi r}=\rho_1 r^{D_{\mathrm{s}}-2}, \tag{14}$$

where $\rho_{\mathrm{L}}(r)$ refers to the density of street distribution, and $\rho_1$ is a proportionality coefficient. Equations (15) is in fact Smeed's density model of transport network (Batty and Longley, 1994; Smeed, 1963). Derivatives of $A(r)$ in equations (10) and $N(r)$ in equations (12) give a density function of traffic nodes in the following form

$$\rho_{\mathrm{N}}(r)=\frac{\mathrm{d}N(r)}{\mathrm{d}A(r)}=\frac{D_{\mathrm{n}}N_1 r^{D_{\mathrm{n}}-1}}{2\pi r}=\rho_1^{*} r^{D_{\mathrm{n}}-2}, \tag{15}$$

where $\rho_{\mathrm{N}}(r)$ denotes the density of node distribution, and $\rho_1^*$ is a proportionality constant. This can be regarded as a generalized form of Smeed's model.

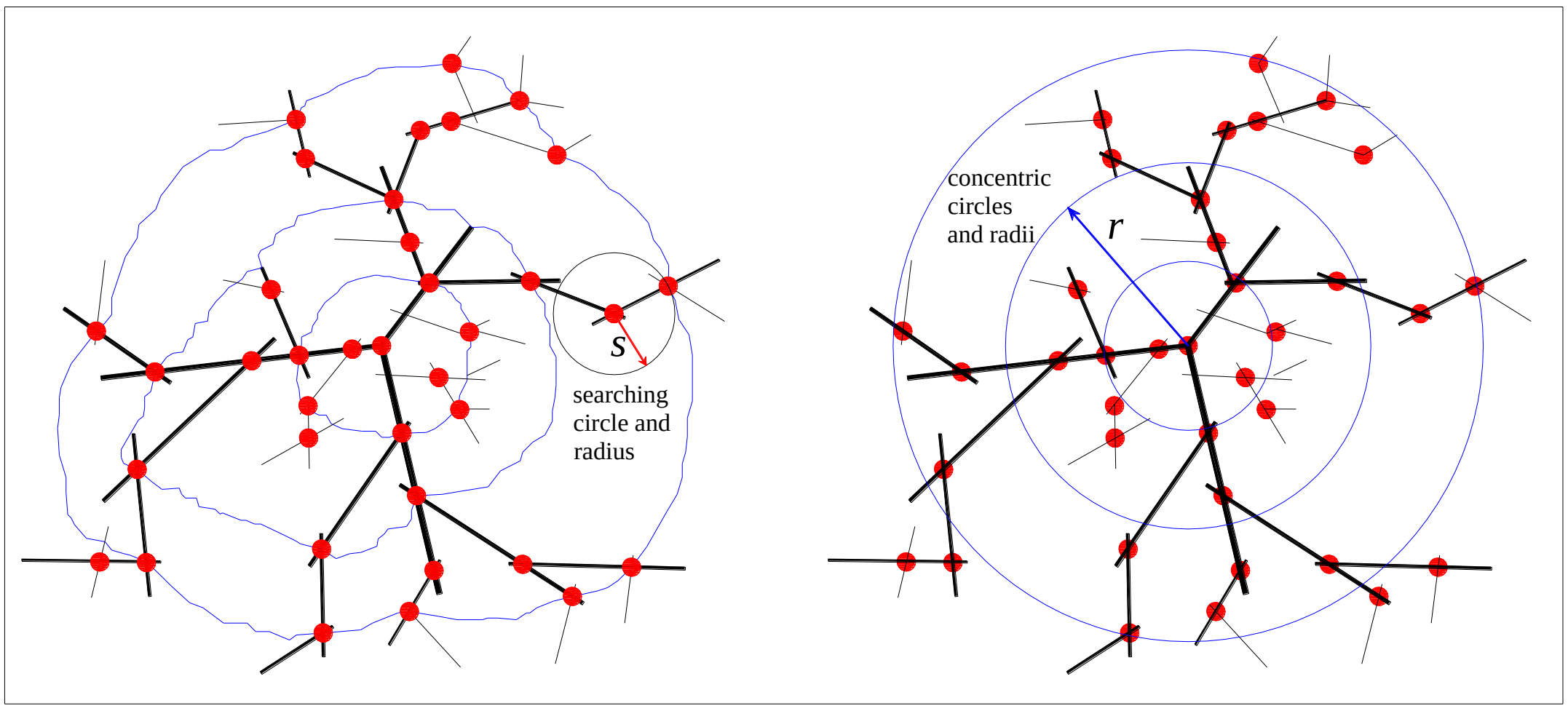

a. Spatial search of urban boundary b. Fractal dimension measurement

**Figure 1 Sketch maps of spatial search of urban boundary and radial fractal dimension measurement of urban traffic networks** [**Note**: The area within an equivalent circle in the second subgraph is equal to the area within the corresponding urban boundary in the first subgraph.]

# 3 Empirical results

## 3.1 Material and methods

As a case study, the models shown above can be applied to the 10 Chinese cities. All these cities are located on the eastern coastal areas of China, including Yangtze River Delta (Shanghai, Nanjing, Hangzhou, Yangzhou, Changzhou, Kunshan), Pearl River Delta (Guangzhou), Shandong Peninsula (Jinan, Weifang), and northeast China (Changchun). The datasets were abstracted from the urban traffic maps of these cities from 2006 year to 2012 year. The maps are made by the cartography sections of local governments. The quality of these maps depends on two factors: traffic route recognition and mapping accuracy requirements. Different cartographers have different recognition patterns of urban streets and roads, which have impact on the mapping effect. Nevertheless, the processes of human cognition for surface features of a geographical region are subject to the same perception rules. The map accuracy rests with confidentiality requirements of the national security department, proportion arrangement of geographical elements, and spatial recognition processes and patterns of cartographers. Despite these defects, it is enough for us to reveal spatial order using these maps. Generally speaking, the accuracy of maps affects the model parameters rather than the mathematical expressions of models. In order to lessen the possible negative influence of map quality, we examined 10 cities, which can be confirmed with one another. If all the datasets from different cities support the same geographical spatial law, the law will be verified and accepted because it is impossible for 10 human settlements to show the same behavior by coincidence. After all, coincidence for many things is an event of small probability.

The data were extracted and processed by ArcGIS technique. First of all, the traffic maps should be digitized in accordance with the following steps. **Step 1**: **map scanning**. The format of digital maps of cities should be transformed into vector graphs, each of which includes a layer of traffic pattern. **Step 2**: **data calibration**. A reference point should be selected inside each urban digital map. The Gauss-Krüger plane rectangular coordinate system was used to calibrate the data in this work. **Step 3**: **map generation**. The method of line breaking can be applied to a digital map for topologic adjustment. By using the mapes, we will extract a dataset of traffic network including a layer of points such as nodes (intersections) and terminals (end points). Then, the ordinary least squares regression can be employed to estimate model parameters. Clauset *et al* (2009) developed new approach based on maximum likelihood method to identify power law and estimate scaling exponent. However, the approach is suitable for binned data but not suitable for our datasets.

Next, a searching radius should be determined to define an urban envelope, i.e., the boundary of a city. Based on an urban envelope, three measurements can be obtained, including urban area inside the urban envelope, node number of traffic network, and total length of transport lines. Changing the searching radius will yield different urban envelopes and the corresponding spatial measurement results. This data extraction can be fulfilled by redevelopment of ArcEngine's function. A computer program of ArcGIS can be written, and a cycle can be designed to control the searching radius. The initial value of the searching radius is $s$=200 meters, which can be termed the minimum radius ($s_{min}$), and the step length of radius increase is $\Delta s$=5 meters. The initial value is empirically chosen by repeated tests. The maximum value of the searching radius ($s_{max}$) depends on urban shape, the spatial pattern of a city's traffic network, and the spatial relationship between a city's figure and map margin. The process of spatial searching of urban boundary starts from a point near the city center with the highest density. The urban area, node number, and street length can be automatically calculated and recorded with the computer program during each searching cycle. Thus a series of variable urban envelopes will form, which look like tree rings (Figure 2). Based on the variable urban boundary lines, three datasets of spatial measurements will yield for each city (see the attached Excel file).

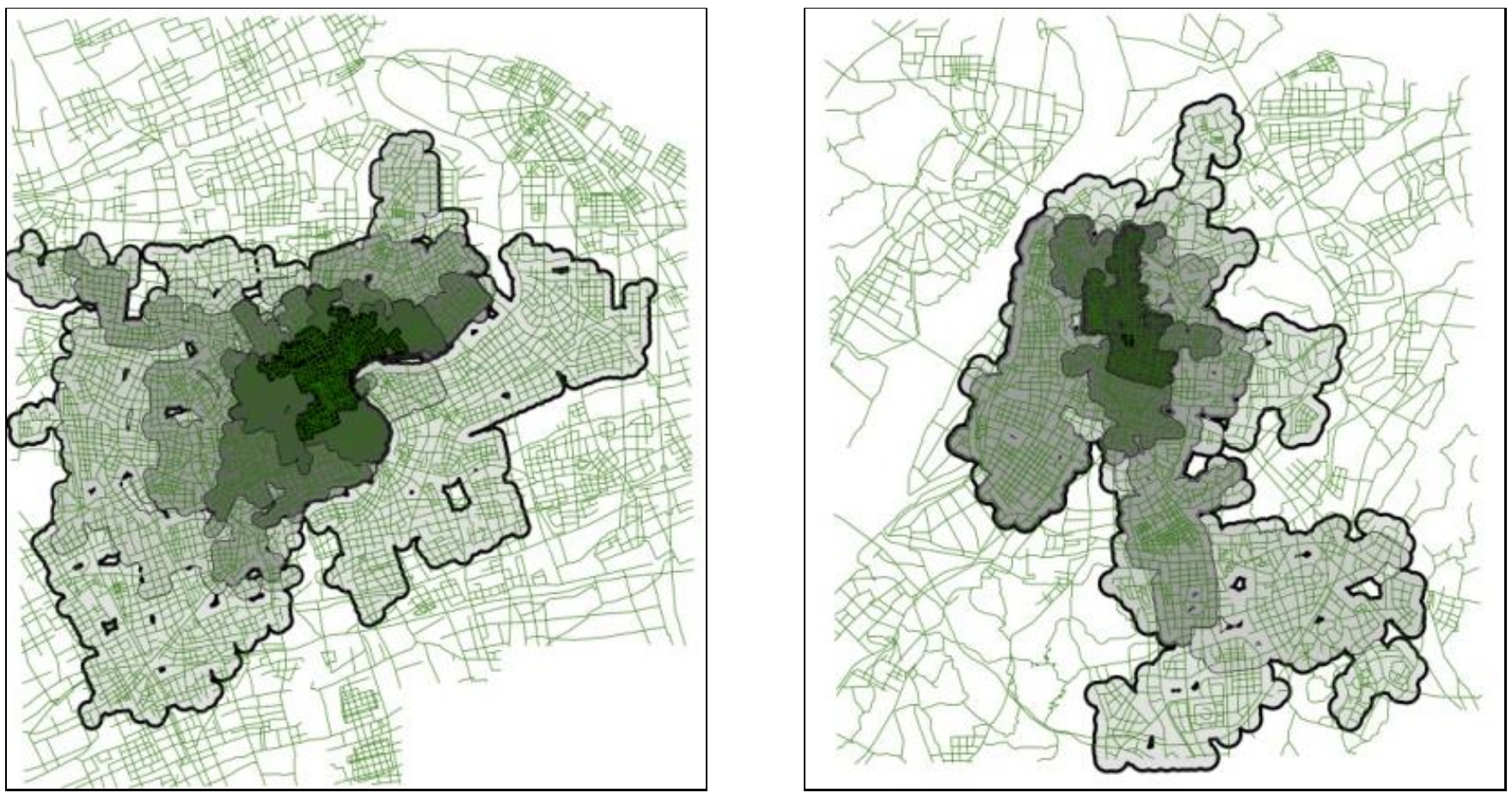

a. Shanghai　　　　　　b. Nanjing

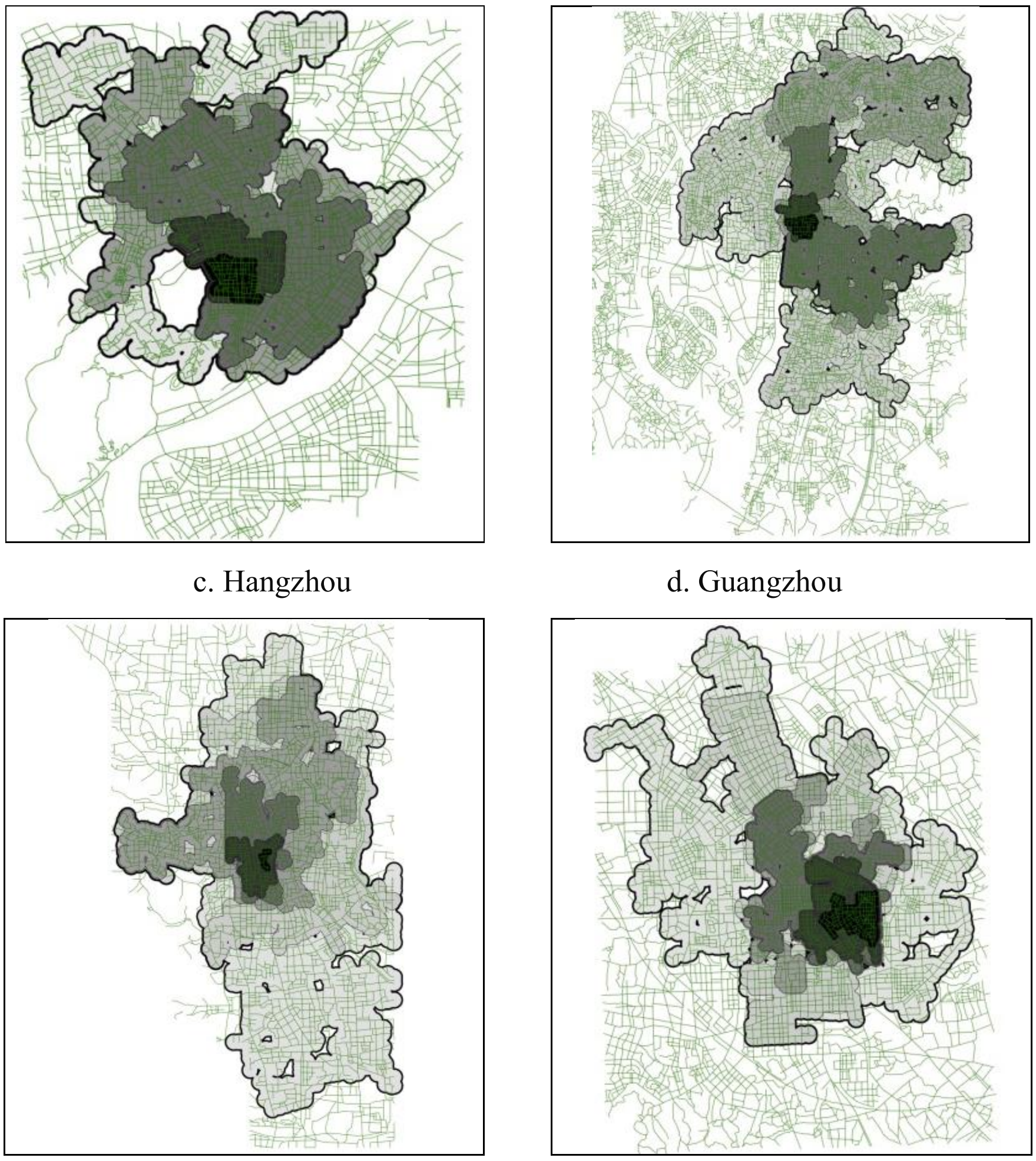

c. Hangzhou d. Guangzhou

e. Jinan f. Changchun

**Figure 1 Six Chinese cities with variable boundaries (2011-2012)** [**Note**: The area of these cities depends on the searching radius. The longer the searching radius is, the larger the urban area will be.]

## 3.2 Results for spatial allometry

Using the datasets of the spatial measurements of the 10 Chinese cities, we can demonstrate the spatial allometric scaling relations between urban area, node number, and street length. Then, we can investigate the relationships between spatial allometry and radial fractal dimension. As indicated above, the geometric measure relations between the three measurements can be formulated with equations (4), (5), and (6). Fitting these equations to the observational data yields allometry models. For example, for Changchun, the provincial capital of Jilin Province in northeast China, the allometric relation between urban area and street length follows a power law. By the least squares computation, we have

$$\hat{L}(s) = 0.2755A(s)^{0.7944}, \quad (16)$$

where the hat "^" suggests that the result is a predicted value (calculated value) rather than an actual observed value (empirical value). The goodness of fit is about $R^2$=0.9987. The scaling exponent is $b$≈0.7944, thus the fractal dimension of the network of streets and roads is estimated as $D_s=2b \approx$ 1.5889. The allometric relation between urban area and node number is as below

$$\hat{N}(s) = 0.0076A(s)^{0.6621}. \quad (17)$$

The coefficient of determination is around $R^2$=0.9980. The scaling exponent is $v \approx$ 0.6621, and the fractal dimension of the node distribution is $D_n=2v\approx$1.3242. The scaling relation between street length and node number is in the following form

$$\hat{N}(s) = 0.0218L(s)^{0.8347}. \quad (18)$$

The squared correlation coefficient is about $R^2$=0.9972. The scaling exponent is $\sigma$≈0.8347, which is the ratio of the node distribution dimension to the network distribution dimension. This fractal dimension ratio can be estimated with equation (9) and the value is $\sigma=v/b=D_n/D_s\approx$ 0.6621/0.7944 ≈0.8334, which is close to the power exponent of equation (18). All the results of parameter estimation are displayed in Table 1. Due to the limitation of the paper space, only the scaling relationships of six cities are shown in Figures 3 and 4 as example.

**Table 1 The allometric scaling exponents, fractal parameters, goodness of fit, and related information of 10 Chinese cities (2006-2012)**

| City | Range of searching radius | Scaling relation | Parameter and statistic: Scaling exponent | Parameter and statistic: Goodness of fit $R^2$ | Fractal parameter |
|---|---|---|---|---|---|
| **Changchun** | 200-440 | $A(s)$-$L(s)$ | $b$=0.7944 | 0.9987 | $D_s$=1.5889 |
| | | $A(s)$-$N(s)$ | $v$=0.6621 | 0.9927 | $D_n$=1.3242 |
| | | $L(s)$-$N(s)$ | $\sigma$=0.8347 | 0.9972 | $D_n/D_s$=0.8334 |
| **Changzhou** | 200-400 | $A(s)$-$L(s)$ | $b$=0.8124 | 0.9990 | $D_s$=1.6248 |
| | | $A(s)$-$N(s)$ | $v$=0.6949 | 0.9976 | $D_n$=1.3898 |
| | | $L(s)$-$N(s)$ | $\sigma$=0.8556 | 0.9991 | $D_n/D_s$=0.8554 |
| **Guangzhou** | 200-345 | $A(s)$-$L(s)$ | $b$=0.9526 | 0.9995 | $D_s$=1.9052 |
| | | $A(s)$-$N(s)$ | $v$=0.9508 | 0.9986 | $D_n$=1.9016 |
| | | $L(s)$-$N(s)$ | $\sigma$=0.9984 | 0.9997 | $D_n/D_s$=0.9981 |
| **Hangzhou** | 200-360 | $A(s)$-$L(s)$ | $b$=0.8740 | 0.9996 | $D_s$=1.7479 |
| | | $A(s)$-$N(s)$ | $v$=0.8034 | 0.9989 | $D_n$=1.6067 |
| | | $L(s)$-$N(s)$ | $\sigma$=0.9194 | 0.9998 | $D_n/D_s$=0.9192 |
| **Jinan** | 280-500 | $A(s)$-$L(s)$ | $b$=0.8620 | 0.9996 | $D_s$=1.7241 |
| | | $A(s)$-$N(s)$ | $v$=0.7669 | 0.9989 | $D_n$=1.5338 |
| | | $L(s)$-$N(s)$ | $\sigma$=0.8898 | 0.9998 | $D_n/D_s$=0.8896 |
| **Kunshan (1)** | 200-500 | $A(s)$-$L(s)$ | $b$=0.7474 | 0.9956 | $D_s$=1.4948 |
| | | $A(s)$-$N(s)$ | $v$=0.5190 | 0.9880 | $D_n$=1.0380 |
| | | $L(s)$-$N(s)$ | $\sigma$=0.6963 | 0.9979 | $D_n/D_s$=0.6944 |
| **Kunshan (2)** | 200-380 | $A(s)$-$L(s)$ | $b$=0.6312 | 0.9982 | $D_s$=1.2625 |
| | | $A(s)$-$N(s)$ | $v$=0.3990 | 0.9840 | $D_n$=0.7981 |
| | | $L(s)$-$N(s)$ | $\sigma$=0.6343 | 0.9923 | $D_n/D_s$=0.6321 |
| | 385-500 | $A(s)$-$L(s)$ | $b$=0.8758 | 0.9998 | $D_s$=1.7515 |
| | | $A(s)$-$N(s)$ | $v$=0.6717 | 0.9990 | $D_n$=1.3433 |
| | | $L(s)$-$N(s)$ | $\sigma$=0.7670 | 0.9993 | $D_n/D_s$=0.7669 |
| **Nanjing** | 275-590 | $A(s)$-$L(s)$ | $b$=0.7958 | 0.9994 | $D_s$=1.5915 |
| | | $A(s)$-$N(s)$ | $v$=0.6716 | 0.9984 | $D_n$=1.3432 |
| | | $L(s)$-$N(s)$ | $\sigma$=0.8442 | 0.9996 | $D_n/D_s$=0.8439 |
| **Shanghai** | 200-450 | $A(s)$-$L(s)$ | $b$=0.7482 | 0.9997 | $D_s$=1.4965 |
| | | $A(s)$-$N(s)$ | $v$=0.5550 | 0.998 | $D_n$=1.1100 |
| | | $L(s)$-$N(s)$ | $\sigma$=0.7420 | 0.9992 | $D_n/D_s$=0.7417 |
| **Weifang** | 385-500 | $A(s)$-$L(s)$ | $b$=0.8842 | 0.9990 | $D_s$=1.7683 |
| | | $A(s)$-$N(s)$ | $v$=0.8373 | 0.9954 | $D_n$=1.6746 |
| | | $L(s)$-$N(s)$ | $\sigma$=0.9480 | 0.9984 | $D_n/D_s$=0.9470 |
| **Yangzhou** | 200-385 | $A(s)$-$L(s)$ | $b$=0.7823 | 0.9991 | $D_s$=1.5646 |
| | | $A(s)$-$N(s)$ | $v$=0.6460 | 0.9983 | $D_n$=1.2920 |
| | | $L(s)$-$N(s)$ | $\sigma$=0.8256 | 0.9988 | $D_n/D_s$=0.8258 |

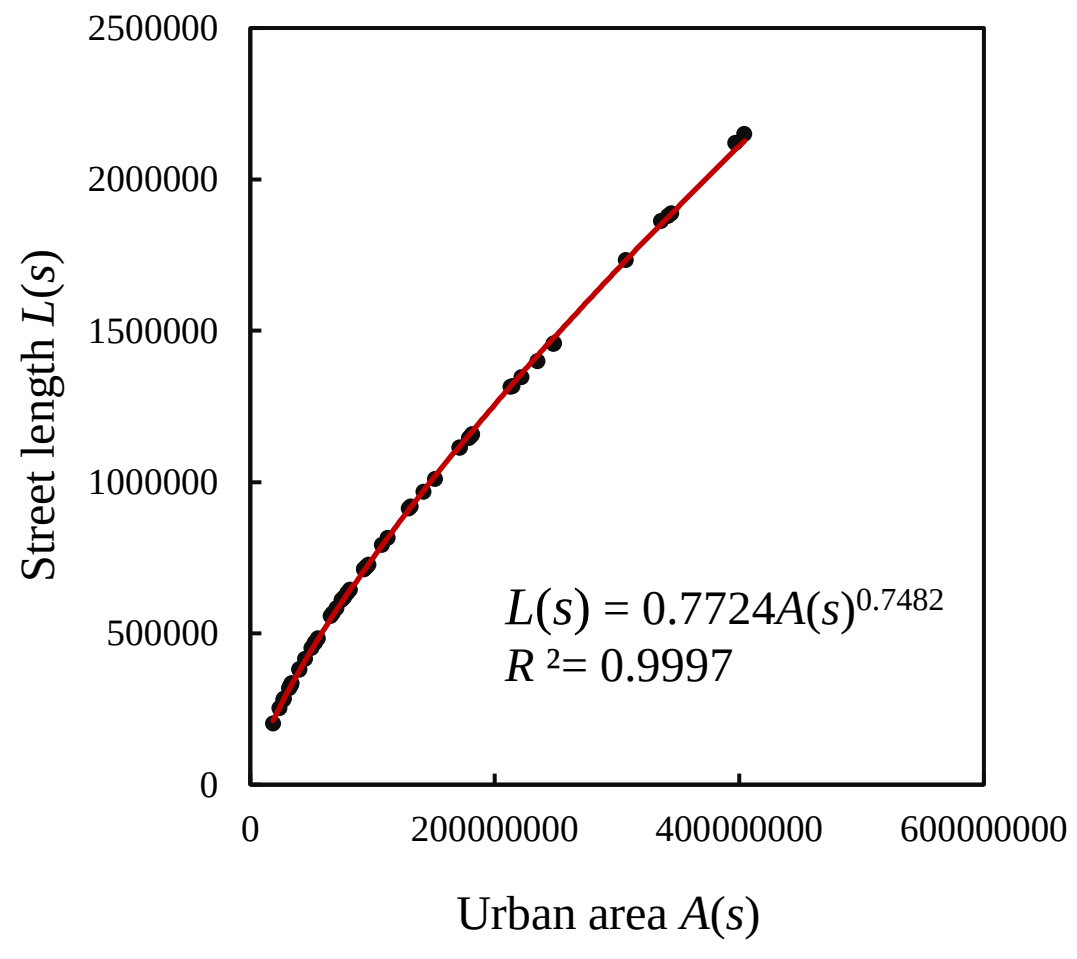

a. Shanghai

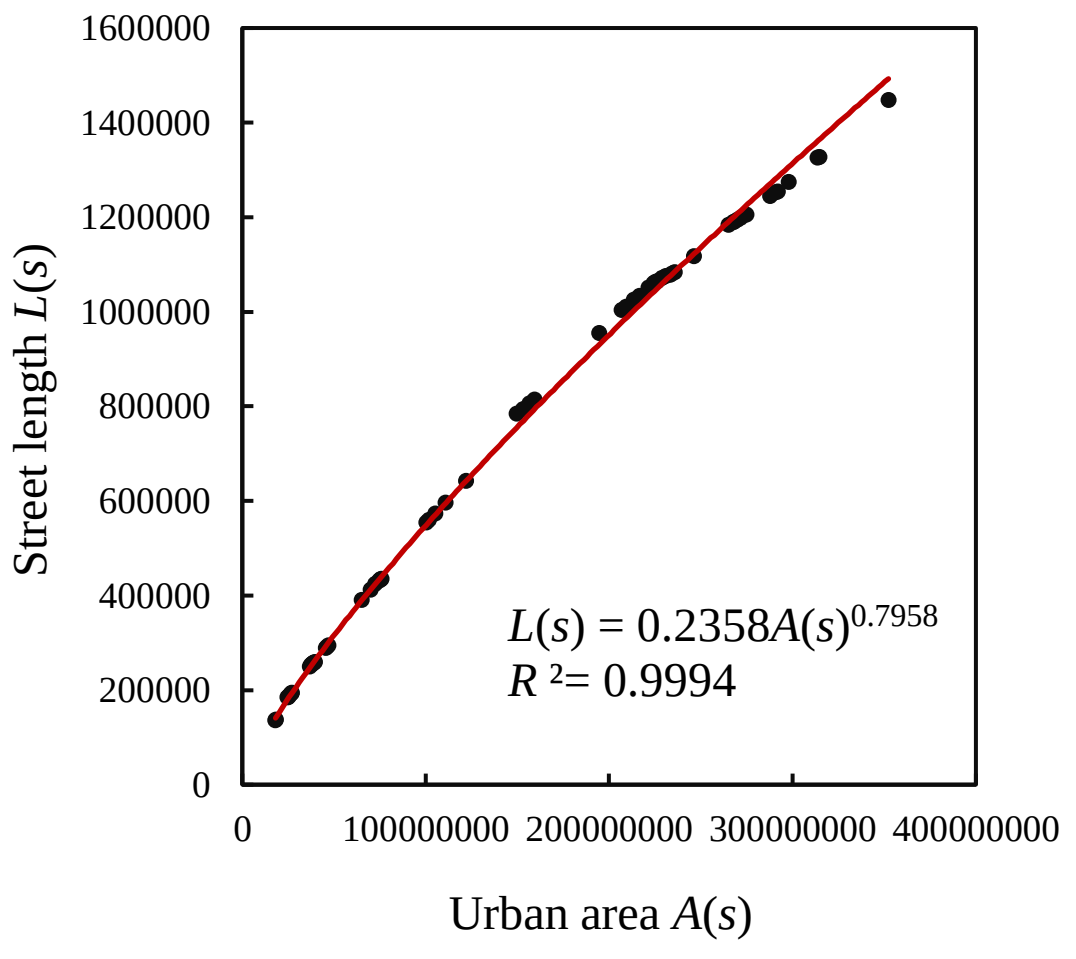

b. Nanjing

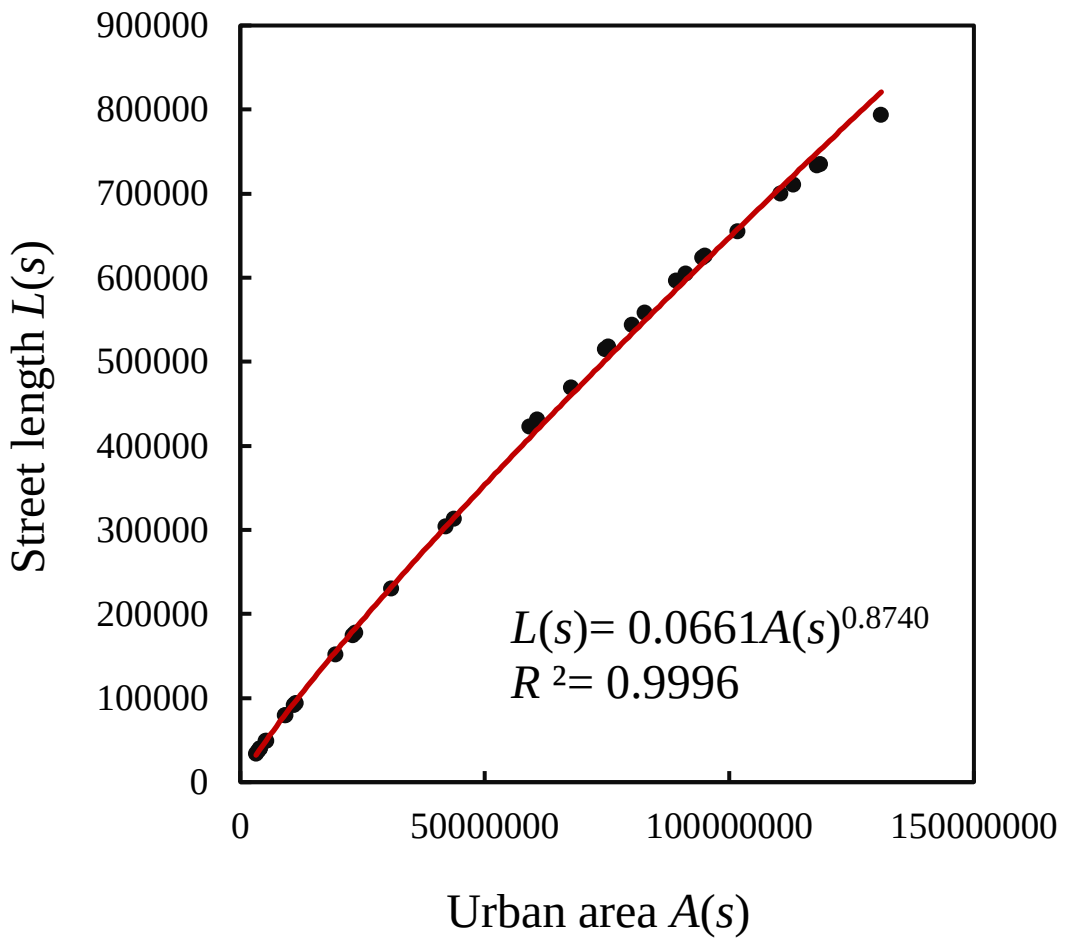

c. Hangzhou

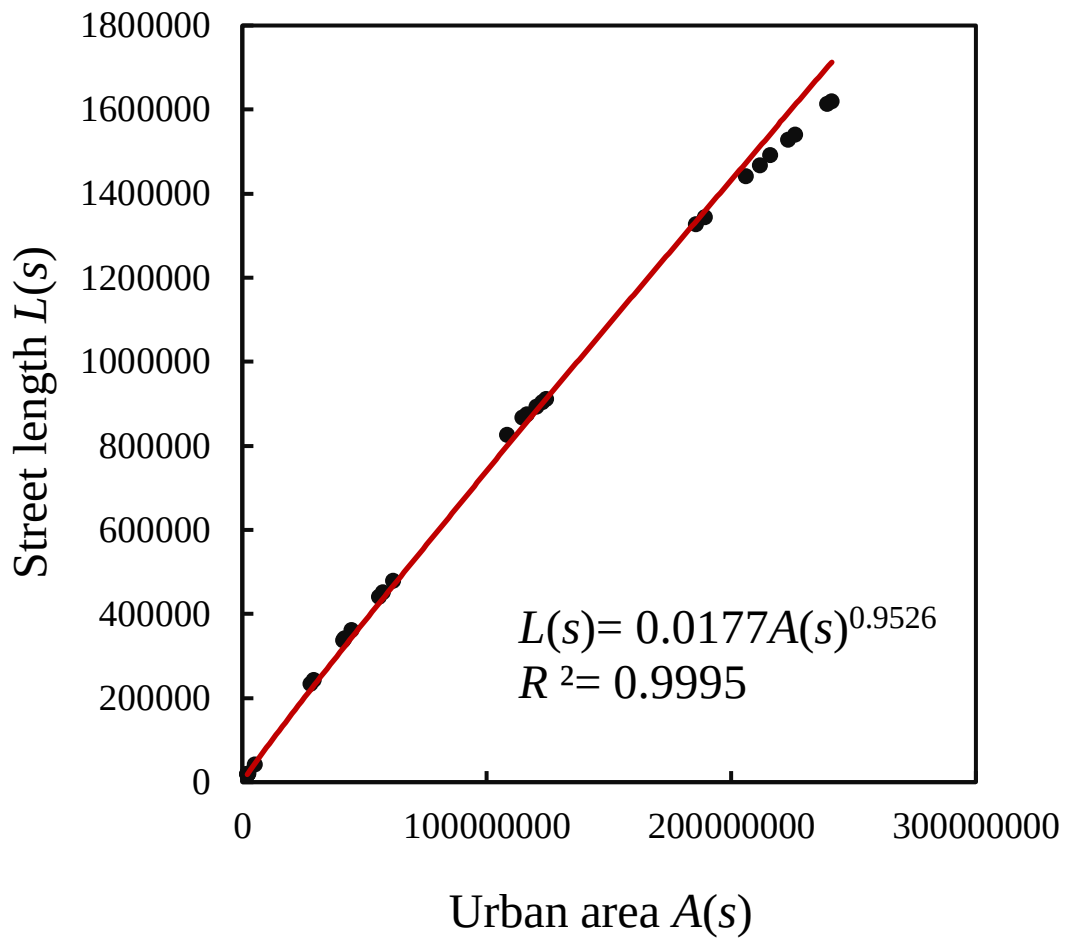

d. Guangzhou

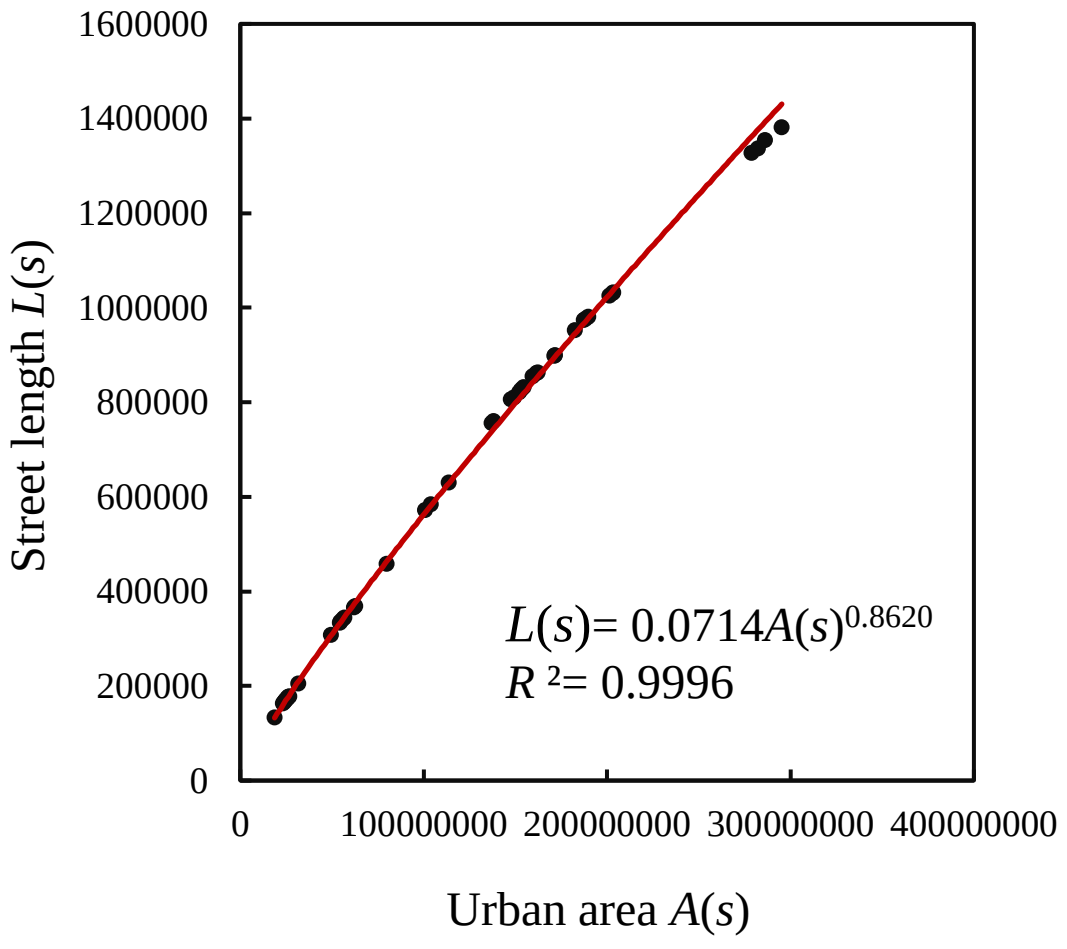

e. Jinan

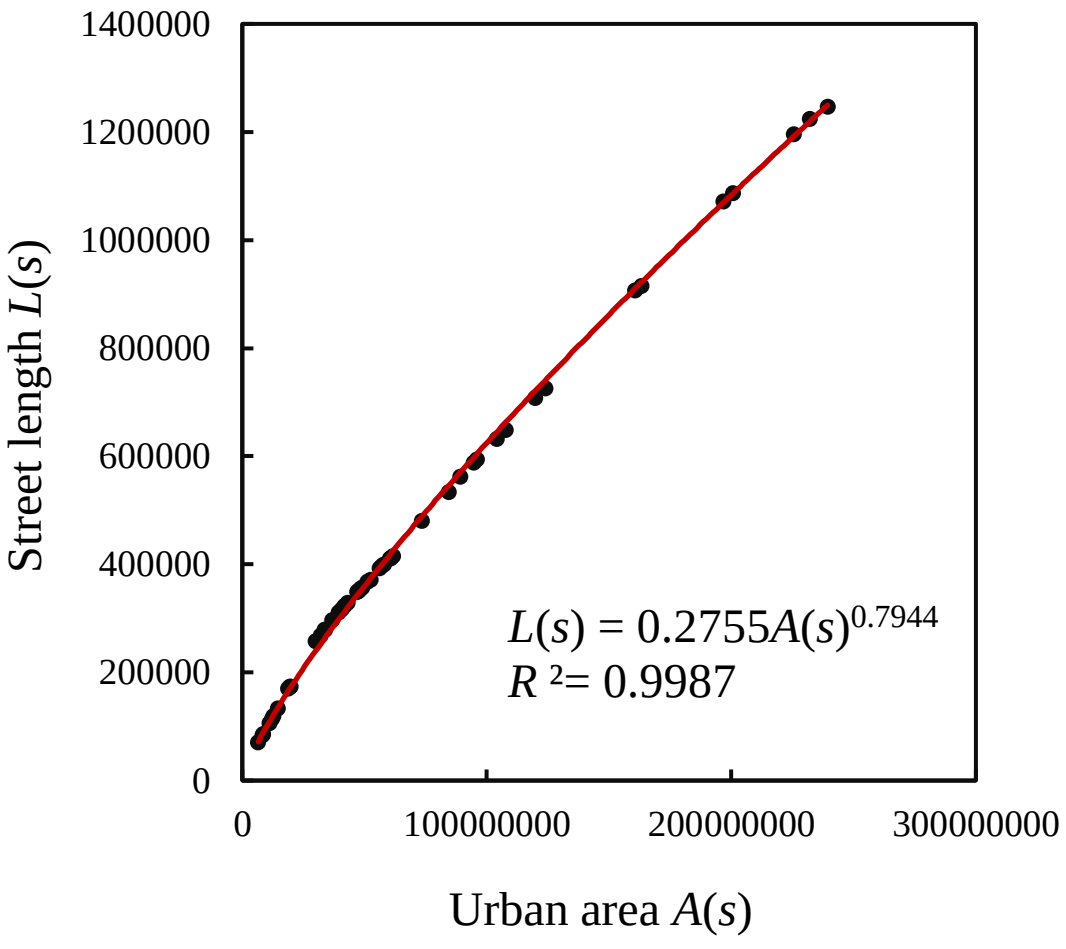

f. Changchun

**Figure 3 The allometric scaling relations between urban area and total street length of six Chinese cities (2011-2012)**

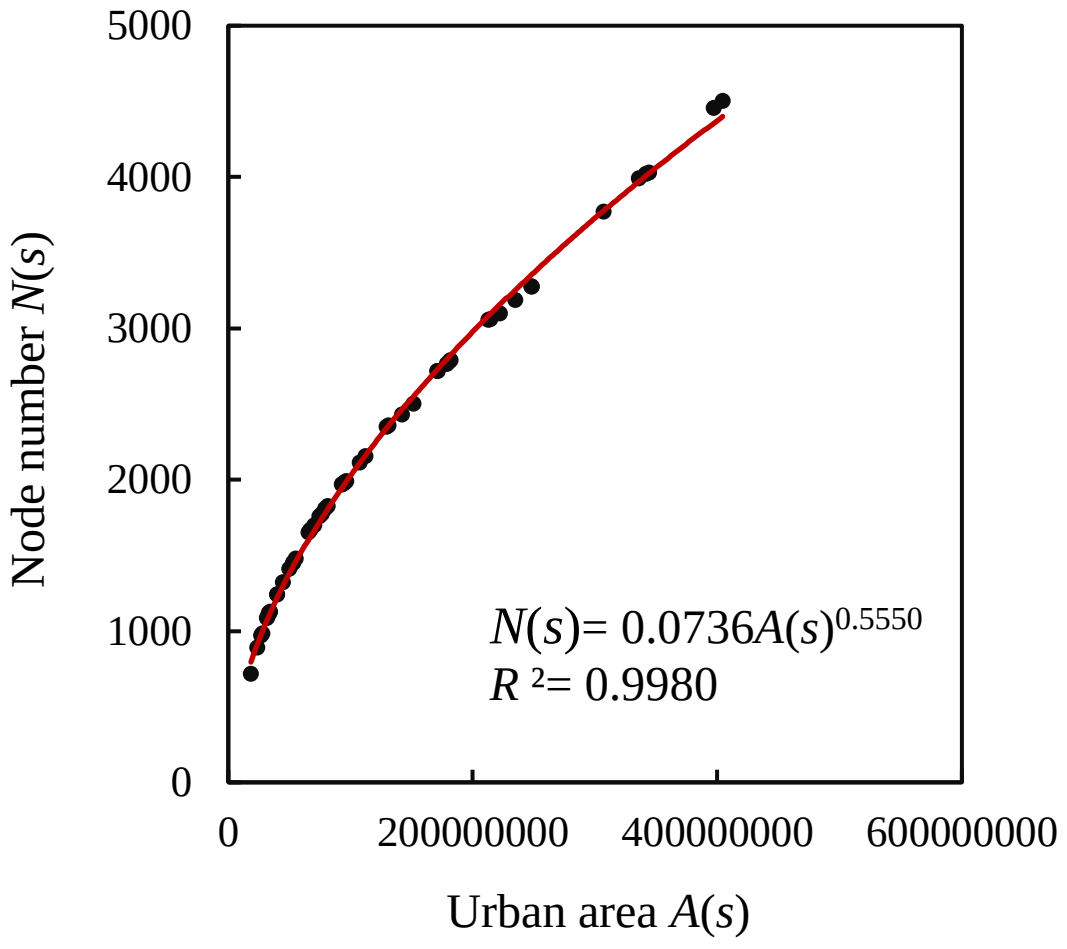

a. Shanghai

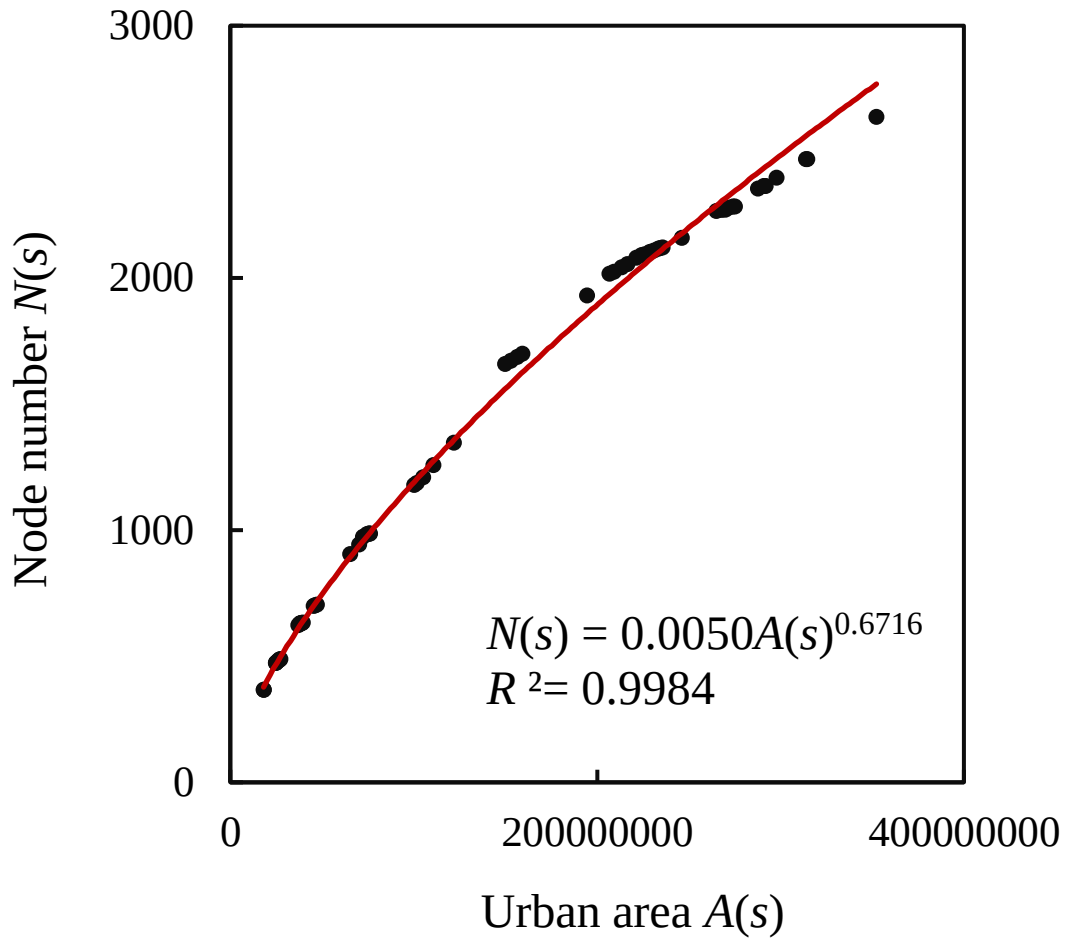

b. Nanjing

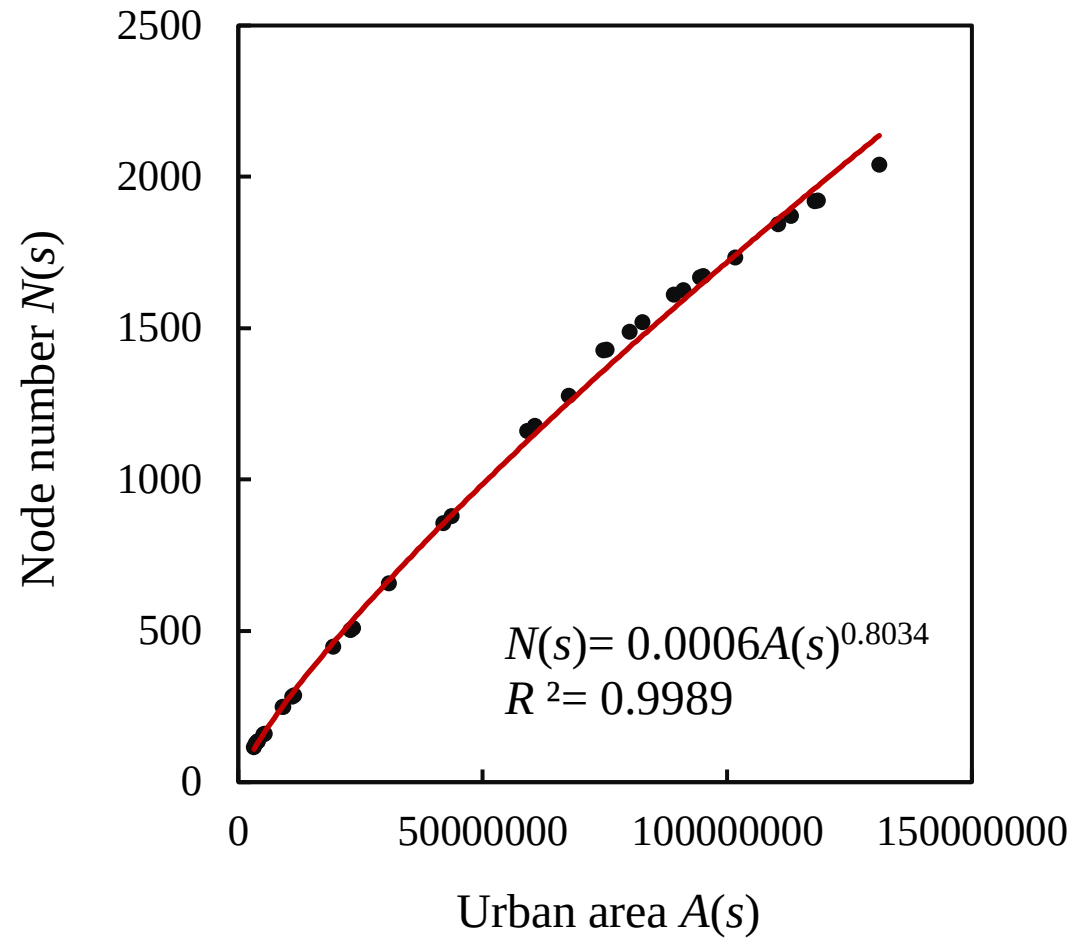

c. Hangzhou

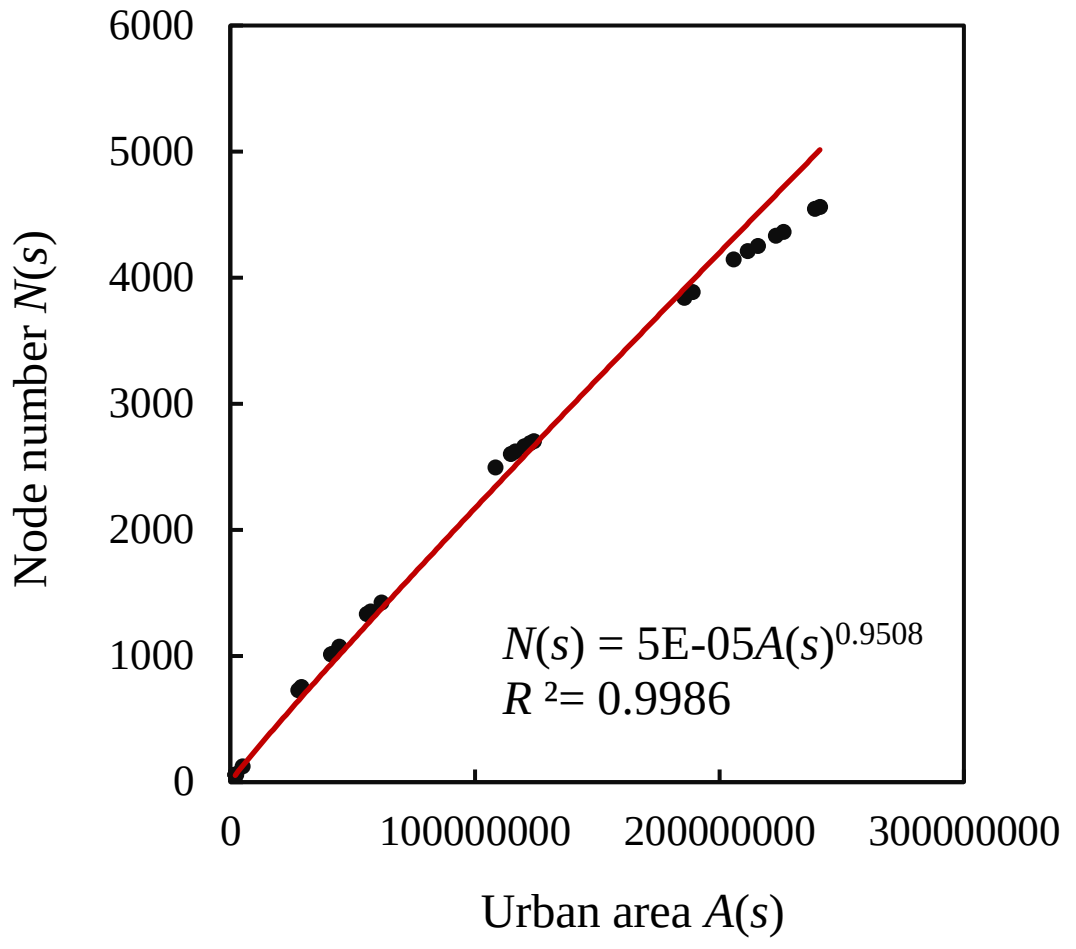

d. Guangzhou

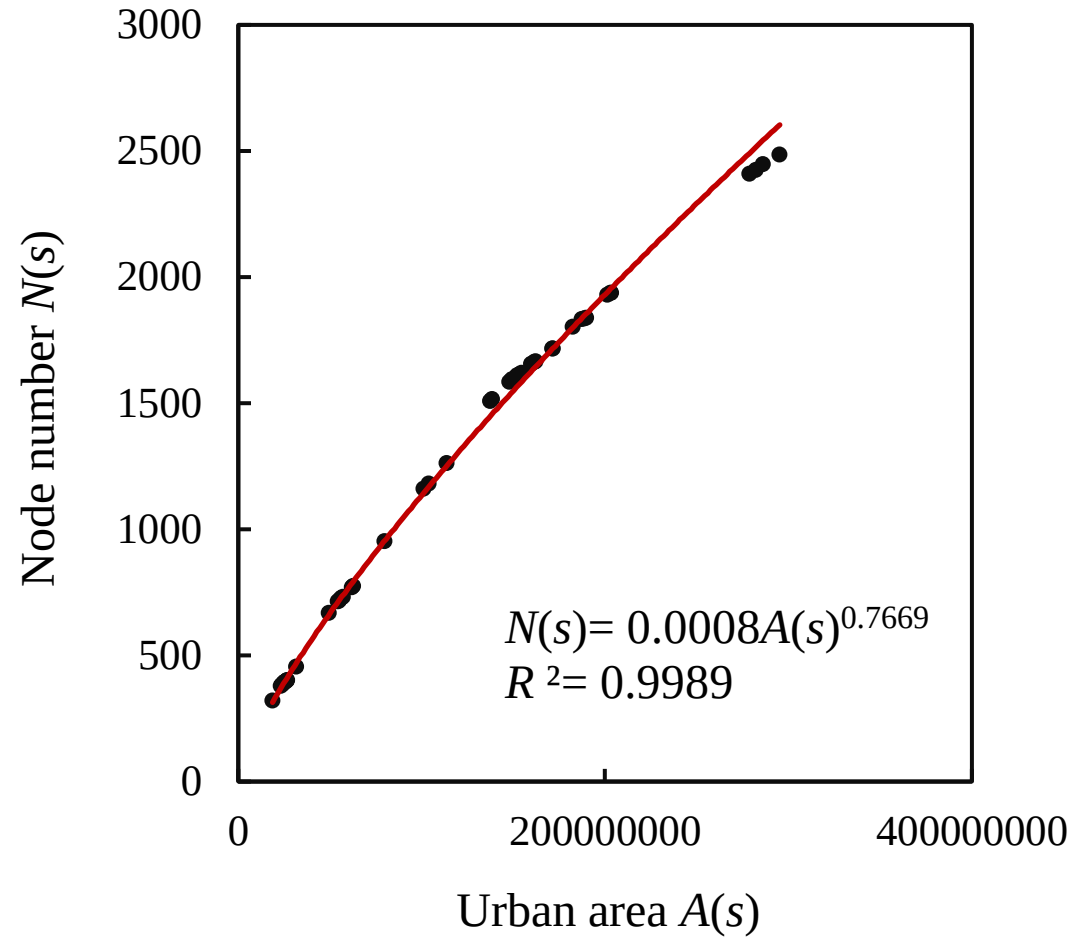

e. Jinan

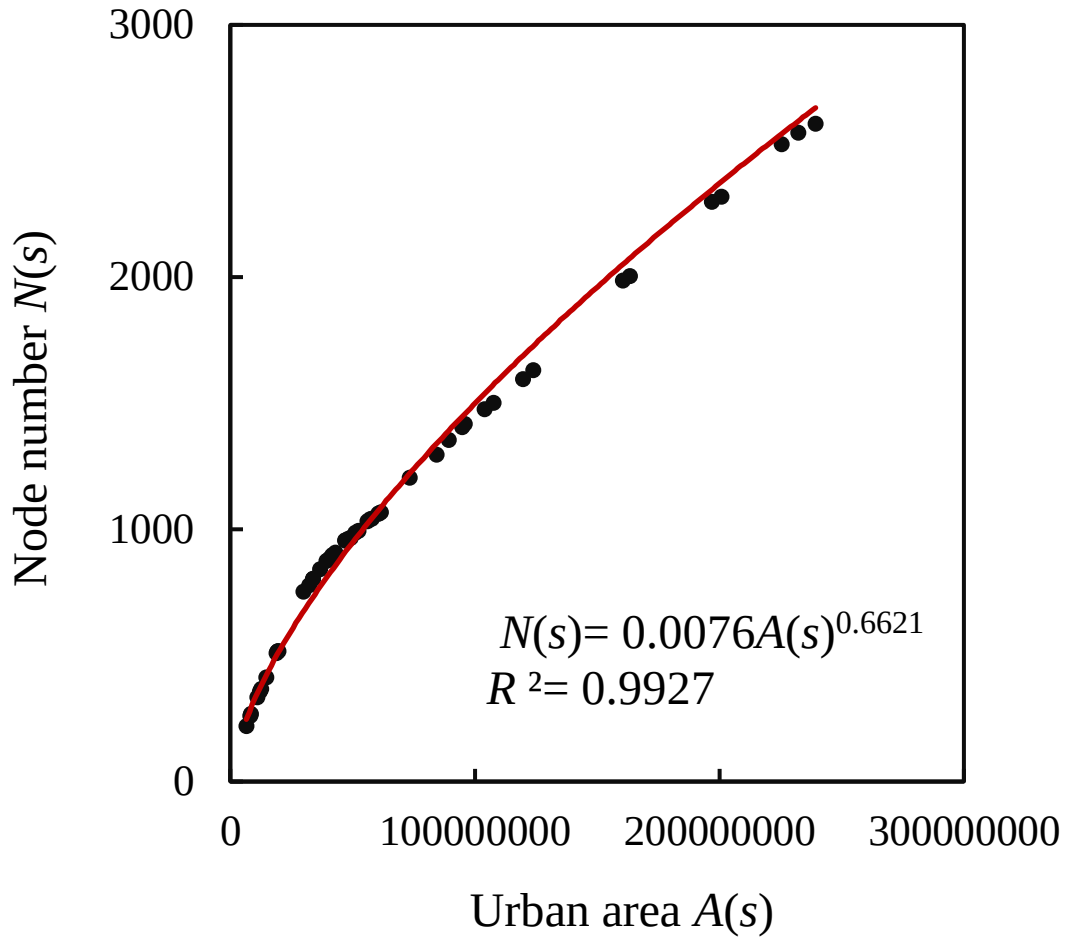

f. Changchun

**Figure 4 The allometric scaling relations between urban area and node number of six Chinese cities (2011-2012)**

The principal points of the empirical data analyses are as follows. **First, all the 10 cities follow the spatial allometric scaling laws.** The numerical relations between urban area, street length, and node number follow power laws to some extent. This suggests that the spatial allometric scaling are universal laws for urban form. **Second, different cities have different scaling range.** The power functions cannot be globally fitted to all the datasets, but can be locally fitted to all the datasets. For example, for the city of Nanjing, the power function cannot be properly fitted to the datasets based on the searching radius from $s_{min}$ ($s$=200) to $s_{max}$ ($s$=590), but if $s$>270, the scaling relations will come into existence. In other words, the scaling range is from $s$=275 to $s$=590, within which the data points can be fitted by a power-law relation. For the city of Kunshan, the scaling breaks and the plot show two scaling ranges. The first scaling range (from $s$=200 to $s$=280) seems to be specious, but the second scaling range (from $s$=285 to $s$=500) is clear and certain (Figure 5). This suggests that the allometric scaling is dynamic evolutional process more than a static spatial pattern. **Third, the line-area scaling and point-line scaling are more significant statistically than the point-area scaling.** Compared with the scaling relation between node number and urban area, the scaling relation between street length and urban area as well as the relation between node number and street length is clearer and shows higher goodness of fit (Figures 3 and 4, Table 1). This suggests that the traffic lines such as streets and roads are more dominated by the scaling law than the nodes. **Fourth, the fractal dimension of lines is higher than that of points.** The empirical data show an inequality such as $D_s$>$D_n$. The fractal dimension is a measure of space filling and reflects urban growth rate (Chen, 2017). The fractal dimension difference between lines and points suggests that the relative growth rate of traffic lines is greater than that of traffic nodes. In short, the traffic lines fill more geographical space than traffic nodes do in given time.

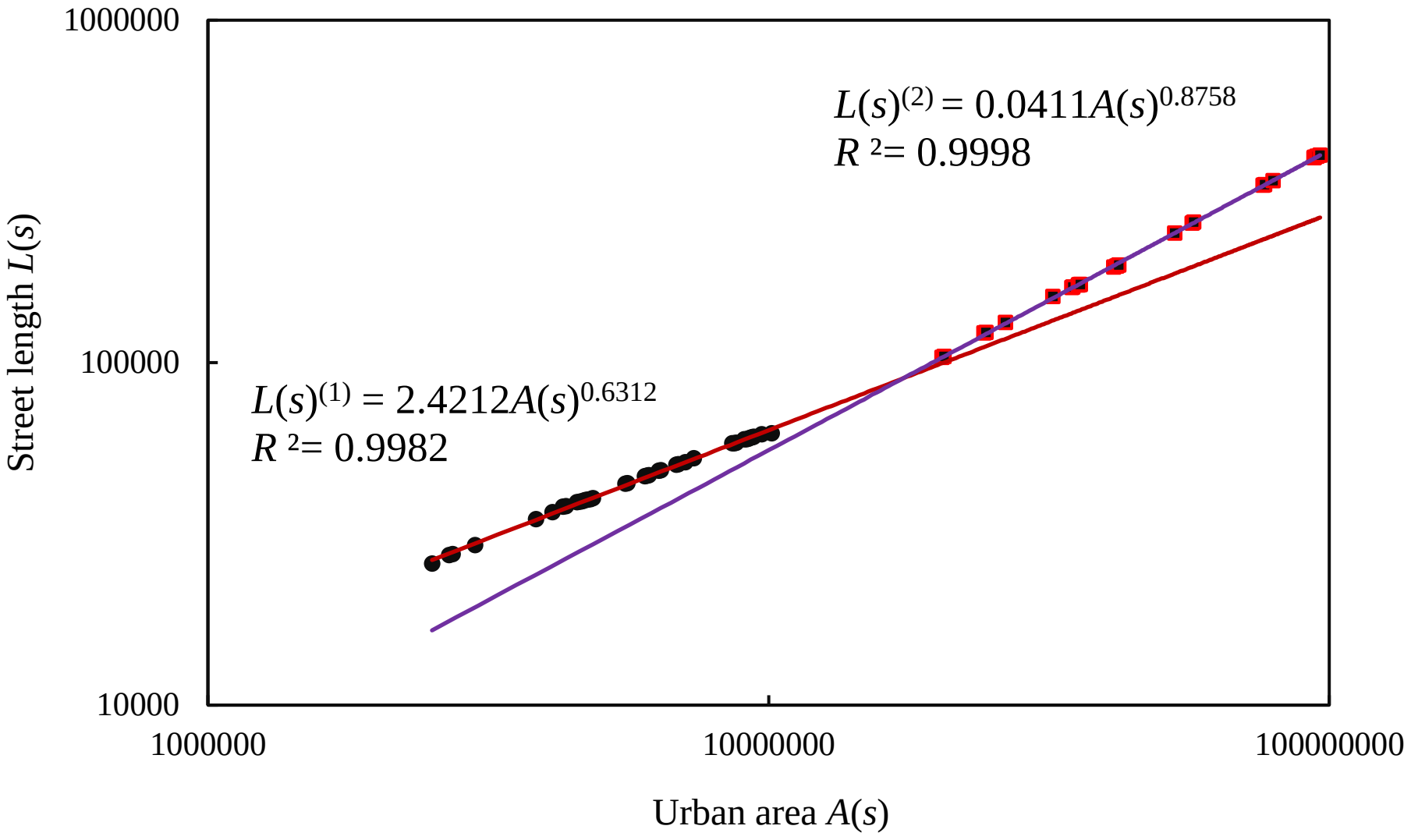

a. Area-length relation

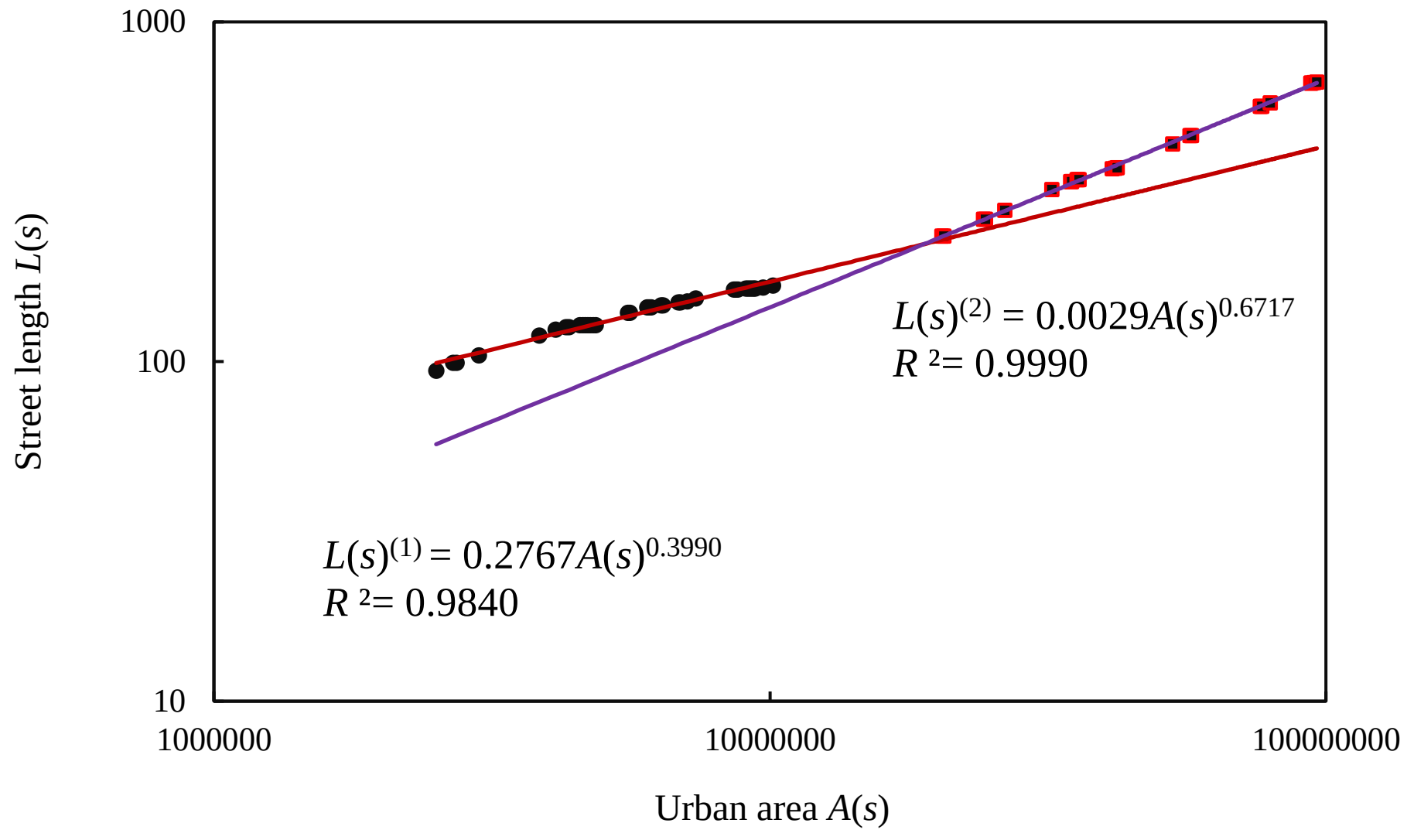

b. Area-number relation

**Figure 5 The patterns of bi-scaling ranges of Kunshan city in log-log plots of spatial allometry**

[**Note**: The superscripts of $L(s)$ in parentheses represent the sequence numbers of scaling ranges. The scaling exponent of the first range is less than that of the second range.]

## 3.3 Results for radial fractal dimension

Based on Smeed's model, the length/number-radius scaling can be employed to estimate radial fractal dimensions of a transport network. Radial dimension can be treated as a kind of fractal parameters reflecting local features of cities (Frankhauser, 1998; Frankhauser and Sadler, 1991). Taking the city of Changchun as an example, we can show how to examine the scaling relation between street length $L(r)$ or node number $N(r)$ and city radius $r$ (Figure 2(a)). The difficult is to determine the center of the city. Two places can be treated as the most possible Changchun's center. One is the old railway station, which is the growth center of the city where its history is concerned; the other is central business district (CBD), which is the developmental center of the city where its present situation is concerned. Fitting equations (11), (12), and (13) to the datasets extracted from the digital map of the city in 2006 yields fractal parameters as follows (Table 2). No matter which place is taken as the measurement center, the fractal character of Changchun's transport network is very significant (Figures 6 and 7). Comparing Table 2 with Table 1 shows that there are similarities and differences between the fractal parameters based on the city radius from an urban center and the estimated values based on the searching radius. The fractal dimension values based on concentric circles are greater than those based on variable urban boundaries. In fact, the spatial search process is a spatial correlation process. The fractal dimension measured by concentric circles corresponds to the capacity dimension, while the fractal dimension measured by variable urban boundaries corresponds to the correlation dimension. In multifractal dimension spectrum, capacity dimension is forever higher than correlation dimension. The spatial allometry is based on variable urban boundary, which is in turn based on CCA. The spatial search of CCA is in fact a process of correlation. Spatial correlation results in scaling behaviors in complex systems such as cities (Chen, 2013; Makse *et al*, 1995; Makse *et al*, 1998). It is hard to make this question clear in a few lines of words, and the relations between two sets of fractal parameters should be specially studied in future.

**Table 2 The fractal parameters of streets and nodes of Changchun's transport network (2006)**

| Measurement center | Scaling relation | Fractal parameter | Standard error | Goodness of fit |
|---|---|---|---|---|
| Railway station | Radius-length | $D_s$=1.8107 | 0.0037 | 0.9995 |
| | Radius-number | $D_n$=1.7823 | 0.0075 | 0.9978 |
| | Length-number | $\sigma^*$=0.9846→$D_n/D_s$ | 0.0027 | 0.9990 |
| CBD | Radius-length | $D_s$=1.8358 | 0.0066 | 0.9984 |
| | Radius-number | $D_n$=1.7462 | 0.0082 | 0.9973 |
| | Length-number | $\sigma^*$=0.9516→$D_n/D_s$ | 0.0017 | 0.9996 |

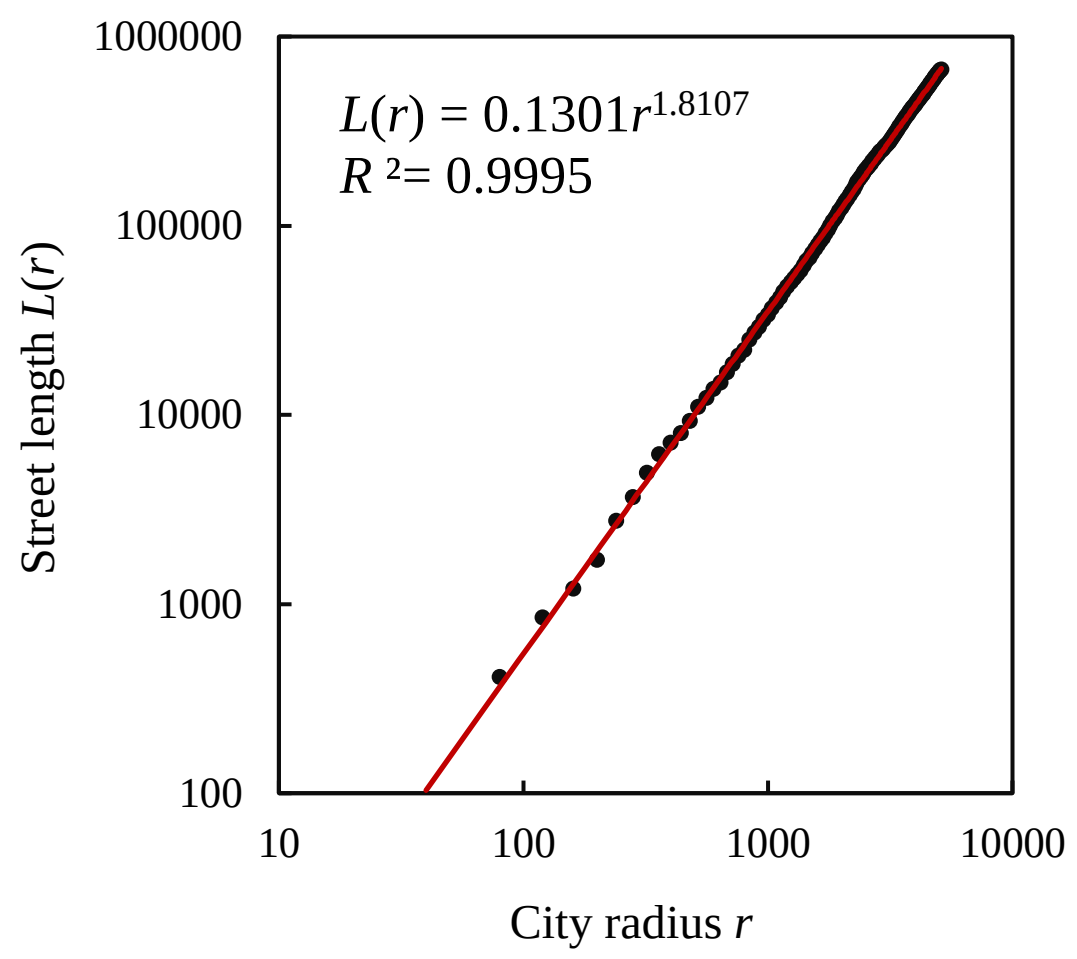

a. Length-Radius

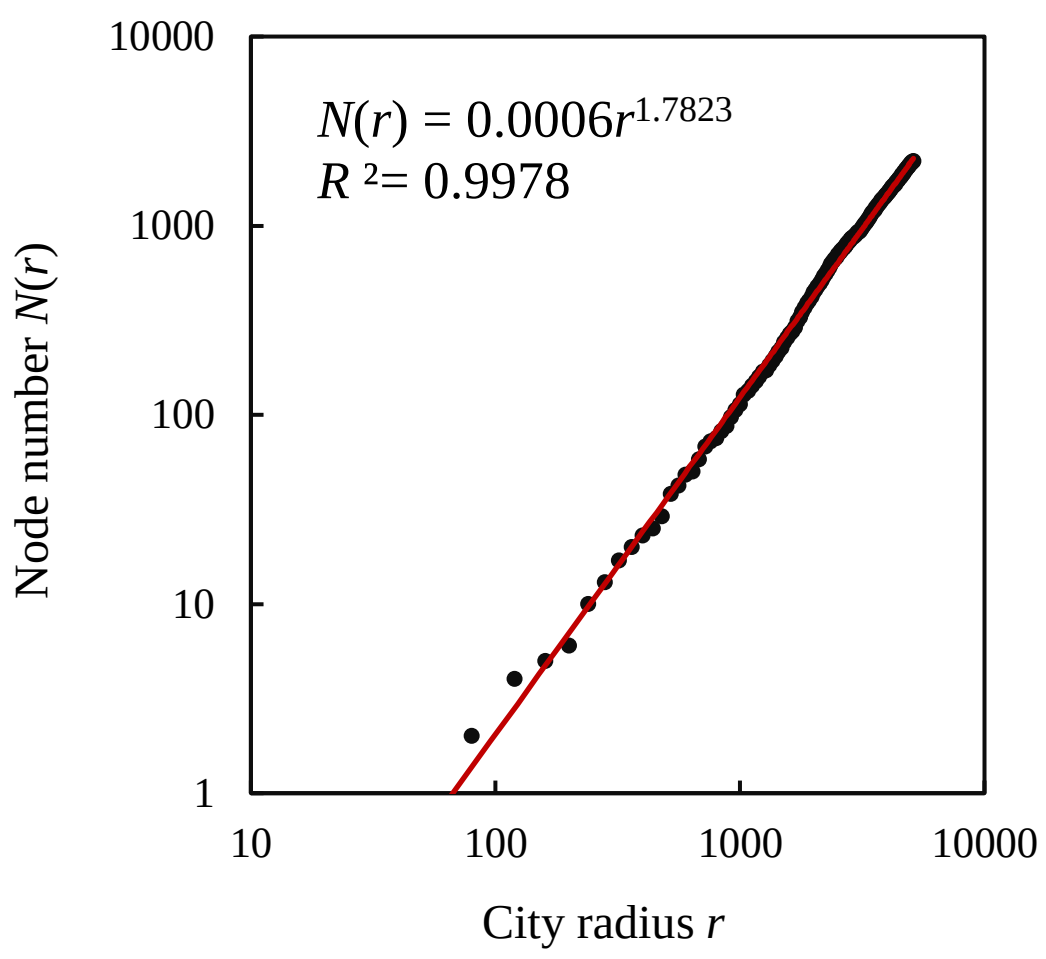

b. Number-Radius

**Figure 6 The fractal dimension measurements of streets and nodes of Changchun's transport network based on railway station (2006)**

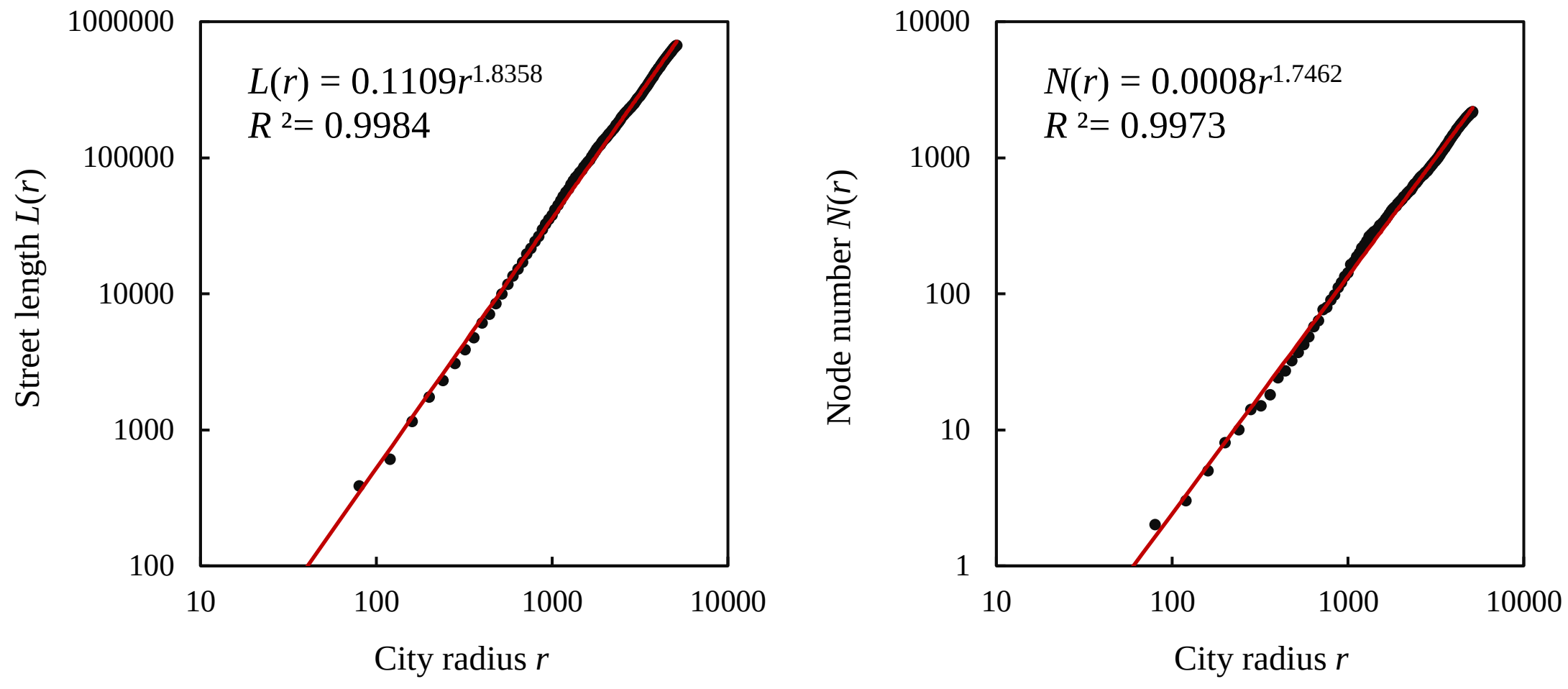

a. Length-Radius b. Number-Radius

**Figure 7 The fractal dimension measurements of streets and nodes of Changchun's transport network based on CBD (2006)**

# 4 Discussion

The improved CCA provides an approach to defining variable urban boundaries, and a set of urban boundary lines can be used to explore the relationships between spatial allometry and city fractals. The *spatial allometry* represents the third allometric scaling in cities. The allometric exponents are the ratios of two types of fractal dimension of urban form. The radial dimension based on concentric circles is to a growing monofractal what capacity dimension is to growing multifractals, while the fractal dimension based on variable urban boundaris is to a growing monofractal what correlation dimension is to growing multifractals. In fact, two kinds of allometric scaling relations between city size and shape have been researched for a long time (Batty and Longley, 1994; Bettencourt, 2013; Lee, 1989; Lo, 2002; Lo and Welch, 1977; Nordbeck, 1971). One is the longitudinal allometry of urban growth, which can be studied using time series data, and the other is the transversal allometry, or cross-sectional allometry, of urban systems, which can be studied using the rank-size series data (Chen, 2014a; Pumain and Moriconi-Ebrard, 1997). The transversal allometry is equivalent to the hierarchical allometry (Chen, 2010). The spatial allometry can be treated the third type of allometry. Compared with the previous studies in literature, this work presents a new allometric relation indeed. Geographical space can be divided into three categories: real space (R-space), phase space (P-space), and Order space (O-space) (Chen, 2014c). The spatial allometry belongs to the R-space, the longitudinal allometry belongs to the P-space, and the transversal allometry and hierarchical allometry belong to the O-space (Table 3).

**Table 3 Three types of geographical space and four kinds of allometric scaling of cities**

| Space | Allometry | Data series | Urban problem | Geography |
|---|---|---|---|---|
| **Real space** | Spatial allometry | Spatial series | Urban form | Pattern |
| **Phase space** | Longitudinal allometry | Time series | Urban growth | Process |
| **Order space** | Cross-sectional allometry | Rank-size series | City size distribution | Hierarchy |
| | Hierarchical allometry | Hierarchical series | Hierarchical structure | Hierarchy |

The spatial allometry models involve three basic geographical elements in a 2-dimensional space, that is, point (traffic node), line (traffic route), and plane (urban region). Based on these spatial elements, we can construct a set of allometric scaling models comprising six fractal measure relations. The six allometric scaling relations form a "matrix", in which the mathematical expressions in the upper triangular matrix are theoretically equivalent to those in the lower triangular matrix because of symmetry (Table 4). In other words, three models are enough to describe the spatial allometry and fractal relations of urban form. In fact, among the three allometry models, one can be derived from other two models. For instance, equation (6) can be derived from equations (4) and (5). This suggests that, in the simplest case, we need only two allometric scaling relations to depict a pattern of urban evolvement. For the power law $y=x^h$ and its inverse function $x=y^g$, theoretically we have a parameter relation such as $h=1/g$, where $x$ and $y$ are variables, $h$ and $g$ refer to power exponents. However, empirically, it can be proved that $h=R^2/g$, where $R^2$ denotes the goodness of fit (Chen, 2017). Therefore, both the models in the upper triangular matrix and those in the lower triangular matrix are useful in practice.

**Table 4 The scaling relations between points, lines, and area and the corresponding fractal parameters**

| | Point (node number) | Line (street length) | Area (urban area) |
|---|---|---|---|
| **Point (node number)** | 1 | $L(s) \propto N(s)^{D_s/D_n}$ | $A(s) \propto N(s)^{D_a/D_n}$ |
| **Line (street length)** | $N(s) \propto L(s)^{D_n/D_s}$ | 1 | $A(s) \propto L(s)^{D_a/D_s}$ |
| **Area (urban area)** | $N(s) \propto A(s)^{D_n/D_a}$ | $L(s) \propto A(s)^{D_s/D_a}$ | 1 |

One of shortcomings in this study is that the relationships between scaling exponent of spatial allometry and radial dimension are not at present derivable from general principles. Maybe it requires much more studies before it will lead us to its underlying rationale. This work is on the basis of the empirical functional relations between three spatial measures, including urban area, road length, and node number and the length of searching radius. From these relations it follows three allometric relations indicative of spatial scale invariance, and these allometric scaling suggests fractal parameters. However, the relations between the three spatial measures and the searching radius are exponential functions indicating characteristic lengths instead of scaling. The spatial allometric scaling is based on variable urban boundaries but goes beyond the length of the searching radii for the boundaries. The reciprocals of the rate parameters, *a*, *c*, and *k*, in equations (1), (2), and (3) are just the characteristic scale parameters. Using the values of $1/a$, $1/c$, and $1/k$ as characteristic lengths, maybe we can define the urban boundary objectively. Of course, this is another topic and should be discussed in a future companion paper.

## 5 Conclusions

The variable urban boundary proved to be related to the scale invariance of urban growth and form. Because of fractality of cities, the boundary of a city defined in a digital map depends on the length of spatial searching radii. Just based on the variable urban boundaries, a set of new allometric scaling relations are found for urban internal structure; based on the spatial allometry, we derive a set of scaling exponents indicative of fractal dimensions. From the theoretical and empirical analyses, the main conclusions of this study can be drawn as follows. **First, a spatial allometric** **scaling exponent are associated with a pair of fractal dimensions.** An allometric exponents are the ratios of one fractal dimension to another fractal dimension of urban structure. Each urban envelope of a city has an equivalent circle. The total length of streets and node numbers of traffic networks within different equivalent circles follow Smeed's law and give a radial dimension of cities. The ratio of fractal dimension of urban nodes to that of urban streets is the scaling exponent of the allometric relation between urban nodes and urban streets. Based on urban area as a 2-dimension measure, the formulae of the fractal dimensions can be derived for streets (lines) and nodes (points) distributions. **Second, the fractal dimension of transport network measured by streets and roads is higher than that of traffic network nodes.** Accordingly, the scaling exponent of the allometric relationship between street lengths and street nodes is less than 1. This suggests that the relative growth rate of streets is greater than that of road nodes, and traffic lines fill more geographical space than traffic points in a city. On the other hand, a set of traffic nodes is in fact a subset of a transport network. In this sense, it suggests that the fractal dimension of a fractal subset is less than that of a fractal set. This can be regarded as a containing principle of fractal dimension of urban form. **Third, an inference is that the spatial allometric scaling yields correlation dimensions.** The fractal dimension values based on variable urban boundaries are lower than the common radial dimension based on concentric circles. The radial dimension seem to correspond to the capacity dimension, while the fractal dimension derived from spatial allometry seems to correspond to correlation dimension in the multifractal spectrum. Correlation dimension is forever less the capacity dimension. Combining the fractal dimension based on concentric circles and those variable urban boundary lines, we will have new radial fractal dimension set for spatial analysis of cities.

### Acknowledgements

This research was sponsored by the National Natural Science Foundation of China (Grant No. 41671167). The supports are gratefully acknowledged.